\documentclass[aps,twocolumn,showpacs]{revtex4}
\usepackage{graphicx}   
\usepackage{latexsym}   

\newcommand{\beq}{\begin{equation}}
\newcommand{\eeq}{\end  {equation}}
\newcommand{\beqar}{\begin{eqnarray}}
\newcommand{\eeqar}{\end  {eqnarray}}

\newcommand{\bold}[1]{\mbox{\boldmath $#1$}}	
\newcommand{\etal}{{\em et al.}}		
\newcommand{\del}{\partial}			
\newcommand{\MeV}{{\rm MeV}}			
\newcommand{\rme}{{\rm e}}			
\newcommand{\grad}{\bold{\nabla}}
\newcommand{\quart}{\mbox{${1\over4}$}}		
\newcommand{\half}{\mbox{${1\over2}$}}		
\newcommand{\cchi}{\bold{\chi}}			

\begin{document}

\title{Fission-fragment mass distributions
from strongly damped shape evolution}

\author{J.~Randrup$^a$, P.~M{\"o}ller$^b$, and A.J.~Sierk$^b$}

\affiliation{
$^a$Nuclear Science Division, Lawrence Berkeley National Laboratory,
Berkeley, California 94720, USA\\
$^b$Theoretical Division, Los Alamos National Laboratory, 
Los Alamos, New Mexico 87545, USA}

\date{July 13, 2011}

\begin{abstract}
Random walks on five-dimensional potential-energy surfaces
were recently found to yield fission-fragment mass distributions 
that are in remarkable agreement with experimental data.
Within the framework of the Smoluchowski equation of motion,
which is appropriate for highly dissipative evolutions,
we discuss the physical justification for that treatment
and investigate the sensitivity of the resulting mass yields
to a variety of model ingredients, including in particular
the dimensionality and discretization of the shape space
and the structure of the dissipation tensor.
The mass yields are found to be relatively robust,
suggesting that the simple random walk presents a useful calculational tool.
Quantitatively refined results can be obtained 
by including physically plausible forms of the dissipation,
which amounts to simulating the Brownian shape motion in an anisotropic medium.
\end{abstract}


\maketitle

\section{Introduction}

A recent study performed random walks on
five-dimensional potential-energy surfaces
and extracted fission-fragment mass distributions 
that are in remarkable agreement with experimental data \cite{RandrupPRL106}.
The striking simplicity of the calculation, its unprecedented predictive power,
and the availability of tabulated potential-energy surfaces
for essentially all nuclei of potential interest \cite{MollerPRC79}
raise the prospect that the method may provide a quantitatively useful
calculational tool for obtaining approximate fission mass yields
for a large region of the nuclear chart.

However, before such applications can be made with any confidence,
a number of issues need to be clarified,
in particular why such a simple, and somewhat arbitrary, treatment
can yield such apparently good results.
Because of the potential utility of the treatment,
we address these issues below.

As discussed already in the pioneering papers by 
Meitner and Frisch \cite{Meitner39} 
and Bohr and Wheeler \cite{BohrNat143,BohrPR56} in 1939,
nuclear fission can be viewed qualitatively as an evolution
of the nuclear shape from that of a single compound nucleus
to two receding fragments.
The character of the shape dynamics is still not well established
but, as a step away from a purely
statistical approach toward a full dynamical treatment,
 it is interesting to explore
scenarios in which the shape evolution is strongly dissipative.

Early studies of fission dynamics with the so-called one-body dissipation
suggested that the nuclear shape motion is strongly damped
\cite{BlockiAP113,SierkPRC21}
and it was advocated that a reasonable starting point for determining
the average evolution would be to balance the conservative force
provided by the potential energy with just the friction force resulting
from the dissipative coupling between the deforming nuclear surface
and the nucleon gas \cite{BlockiAP113,RandrupAP125}.
A simple feature of strongly damped motion is that
the inertial forces are relatively unimportant 
so it is less crucial to know the inertial-mass tensor for the shape motion. 

\phantom{phantom text}

The earliest numerical studies of dissipation in fission dynamics were
focused on the damping effect on the mean motion only, using various
physical models for dissipation, including inertial effects, and using
macroscopic potentials \cite{DaviesPRC13,SierkPRC17,BlockiAP113,SierkPRC21}.
Consideration of the stochastic force in fission dynamics began
as early as 1940, when Kramers considered the average delay in
establishing a stationary flow rate over a one-dimensional barrier,
thus inferring an increase of the fission lifetime due to dissipation
\cite{KramersPhys7}. Further approximate treatments of the Fokker-Planck
equation in one or two dimensions began around 1980 
\cite{GrangePL88B,HofmannPL122B,GrangePRC27,NixNPA424,ScheuterPL149B}.
These calculations often retained the assumption of a constant inertia
and dissipation, with very simplified potentials.
About a decade later, numerical investigations of Langevin equations
for dynamics including inertia, damping, and Markovian stochastic forces
were begun by several groups; reviews of such work were given in 
Refs.\ \cite{AbePhysRep275,FrobrichPhysRep292}.
These types of simulations continue 
\cite{KarpovPRC63,ChaudhuriPRC63,NadtochyPRC75}. 
They employ two or three shape degrees of freedom, macroscopic
potential energies, fluid dynamic inertias, and more recent calculations
usually use some form of one-body dissipation. Because of the use of
macroscopic energies, they are often explictly characterized as applying
to systems with significant excitation energy.

In contrast,
the present investigation focuses on systems having relatively low
excitation energy, such as those produced by thermal neutrons,
where it is essential to include microscopic (shell) effects 
in the potential energy.
For 5,254 even-even nuclei with $170\leq A\leq330$,
such potential energies have been calculated 
with the macroscopic-microscopic method 
on a five-dimensional lattice of 5,315,625 shapes \cite{MollerPRC79}.
These potential-energy surfaces are the most comprehensive
currently available and form the basis for our present studies.

~\vfill

\section{Formal framework}
\label{frame}

We picture the nuclear fission process as an evolution of the nuclear shape 
from a relatively compact mononucleus to a dinuclear configuration.
The nuclear shape is described by a set of parameters, $\cchi=\{\chi_i\}$,
whose time development is the result of a complicated interplay between
a variety of effects, as we now discuss.

Most basic, and also most easy to understand, 
is the potential energy associated with a given shape, $U(\cchi)$,
for which a number of relatively mature models have been developed.
We employ potential-energy surfaces that have been calculated
with the macroscopic-microscopic method
in which the potential energy is the sum of shape-dependent
macroscopic (liquid-drop type) terms and a microscopic correction 
that reflects the structure of the single-particle levels
in the effective potential associated with the specified nuclear shape
\cite{MollerNat409}.
The potential energy provides the driving force, $\bold{F}^{\rm pot}(\cchi)$,
which has the components $F_i^{\rm pot}=-\del U/\del\chi_i$.

The driving force seeks to change the nuclear shape and 
the associated matter rearrangement gives rise to a collective kinetic energy.
Furthermore, the shape degrees of freedom are coupled dissipatively
to the internal nuclear degrees of freedom and, as a result,
the shape evolution is both damped and diffusive.
The nuclear shape dynamics should therefore be treated by transport methods 
that allow for the stochastic elements of the dynamics.
Stochastic transport approaches to nuclear dynamics have been 
reviewed in Refs.\ \cite{AbePhysRep275,FrobrichPhysRep292}.

The Lagrangian associated with the shape degrees of freedom
has a standard form,
\beq
{\cal L}(\dot{\cchi},\cchi)\ = \
\half\sum_{ij}\dot{\chi}_i\, M_{ij}(\cchi)\,\dot{\chi}_j - U(\cchi)\ ,
\eeq
and the (generalized) momentum can be obtained as
$\bold{\pi}\equiv\del{\cal L}/\del\dot{\cchi}$,
with the components being $\pi_i=\sum_jM_{ij}\dot{\chi}_j$.
The inertial-mass tensor, $\bold{M}(\cchi)$,
is not as well understood as the potential energy
and cannot yet be calculated with comparable accuracy.
In the present investigation, the inertia is ignored 
because it is expected to play a relatively minor role
if the dissipation is strong (see below).

To a first approximation, the average effect of the coupling between
the shape and the residual system is a simple friction
described by a Rayleigh dissipation function,
\beq
{\cal F}(\dot{\cchi},\cchi)\ =\ 
\half\sum_{ij}\dot{\chi}_i\,\gamma_{ij}\,(\cchi)\,\dot{\chi}_j\ 
=\	\half\dot{Q}\ ,
\eeq
which is equal to half the average rate at which energy is being transferred
to the internal degrees of freedom.
The dissipation tensor $\bold{\gamma}(\cchi)$ governs
the associated friction force $\bold{F}^{\rm fric}(\cchi)$, with
$F_i^{\rm fric}=-\del{\cal F}/\del\dot{\chi}_i=-\sum_j\gamma_{ij}\dot{\chi}_j$.
We shall invoke the simple {\em wall formula} \cite{BlockiAP113}
to obtain estimates of $\bold{\gamma}(\cchi)$ (see Sect.\ \ref{mu}).

The average shape evolution $\bar{\cchi}(t)$
is then governed by the Lagrange-Rayleigh equation,
\beq\label{L-R}
{d\over dt}\bar{\bold{\pi}}\ =\
{\del{\cal L}\over\del\cchi}-{\del{\cal F}\over\del\dot{\cchi}}\ ,
\eeq
where the derivatives should be evaluated along the mean trajectory.
When the damping is strong, the resulting motion is slow.
In that scenario, the acceleration terms
as well as terms of second order in the velocities may be neglected.
The corresponding creeping evolution is then determined by the demand
that the friction force exactly counter balance the driving force,
i.e.\ $\bold{F}^{\rm pot}+\bold{F}^{\rm fric}\doteq\bold{0}$.
This equation can then be solved for the average velocity 
$\dot{\overline{\cchi}}$,
$\dot{\overline{\chi}}_i \doteq \sum_j\mu_{ij}F^{\rm pot}_j$,
where the {mobility tensor} \bold{\mu}\
is the inverse of the dissipation tensor \bold{\gamma}.
Because the dissipation rate $\dot{Q}$ is always positive,
the tensor \bold{\gamma} is positive definite; so its inverse \bold{\mu}
always exists and its eigenvalues are positive.

The friction force represents the average of the interactions
of the shape with the internal degrees of freedom.
As is common, we shall assume that the remaining interaction
is stochastic and the associated force is denoted by $\bold{F}^{\rm ran}(t)$.
(Its time dependence is indicated explicitly because it is expected 
to vary rapidly on the time scale of the shape evolution.)
By definition, it vanishes on the average,
$\langle\bold{F}^{\rm ran}(t)\rangle=\bold{0}$,
and we assume that its time dependence is Markovian,
so $\langle{F}_i^{\rm ran}(t){F}_j^{\rm ran}(t')\rangle
=2T\gamma_{ij}\delta(t-t')$, 
where $T(\cchi)$ is the shape-dependent nuclear temperature (see later).

The actual shape evolution is thus both damped and diffusive and
the trajectory $\cchi(t)$ is governed by the Smoluchowski equation of motion
in which the driving force from the potential is
counterbalanced by the full dissipative force, i.e.\ 
$\bold{F}^{\rm pot}+\bold{F}^{\rm fric}+\bold{F}^{\rm ran}\doteq\bold{0}$.
This condition immediately yields the instantaneous velocity,
\beq\label{EoM}
\dot{\cchi}(t)\ \doteq\
\bold{\mu}\cdot(\bold{F}^{\rm pot}+\bold{F}^{\rm ran}(t))\ .
\eeq
We assume that the nuclear shape evolution is described by
this equation and it is thus akin to Brownian motion.
The net displacement accumulated in the course of a brief time interval
$\Delta t$ is then
\beq\label{dchi}
\delta\cchi =  \int_0^{\Delta t}\!\dot{\cchi}(t)\,dt
=	\bold{\mu}\cdot\!\left[\bold{F}^{\rm pot}\Delta t 
+	\!\int_0^{\Delta t}\!\bold{F}^{\rm ran}(t)\,dt\right] ,
\eeq
where we have chosen $\Delta t$ to be so small that both the driving force
and the mobility tensor can be considered as constant.
The first term in the above expression (\ref{dchi})
is deterministic and represents the average displacement,
corresponding to the mean trajectory provided by the
Lagrange-Rayleigh equation (\ref{L-R}),
whereas the second term is stochastic
and arises from the inherent thermal fluctuations
that give the evolution a diffusive character.

Because of the diffusive nature of the dynamics,
it is appropriate to describe the system by a probability density
$P(\cchi;t)$.
The average shape is then described by
\beq
\bar{\cchi}(t)\ =\ \int\, d\cchi\, \cchi P(\cchi;t)\ ,
\eeq
while the fluctuations around the average are characterized
by the correlation tensor $\bold{\sigma}(t)$ having the elements
\beq
\sigma_{ij}(t)\ =\ 
\int d\cchi\,(\chi_i-\bar{\chi}_i(t))\,(\chi_j-\bar{\chi}_j(t))\,P(\cchi;t)\ .
\eeq

When starting from a given shape specified by $\cchi_0$,
we have $P(\cchi;t=0)=\delta(\cchi-\cchi_0)$
and the distribution then shifts and broadens in the course of time.
In the Fokker-Planck approximation, this evolution is described by
\beq\label{FPE}
{\del\over\del t}P\ =\ \left[-\sum_i{\del\over\del\chi_i}{\cal V}_i +
\sum_{ij}{\del\over\del\chi_i}{\del\over\del\chi_j}{\cal D}_{ij}\right]
P(\cchi;t)\ ,
\eeq
where $\bold{\cal V}(\cchi)$ is the drift coefficient (vector)
and $\bold{\cal D}(\cchi)$ is the diffusion coefficient (tensor).
These transport coefficients are simply related to the
early evolution of an initially narrowly defined distribution, namely
\beq
{d\over dt}\bar{\cchi} = \bold{\cal V} = \bold{\mu}\cdot\bold{F}^{\rm pot}\ 
,\,\,\ {d\over dt}\bold{\sigma} = 2\bold{\cal D} = 2\bold{\mu}T\ .
\eeq
The drift rate $\dot{\bar{\cchi}}$ follows immediately from (\ref{dchi})
and the diffusion rate $\dot{\bold{\sigma}}$ is also readily obtained
because the covariance matrix for the changes in \cchi\ is given by
\beqar
\langle\delta\cchi\delta\cchi\rangle\!
&=&\!	\int_0^{\Delta t}\!\int_0^{\Delta t}\!
	\langle\dot{\cchi}(t)\,\dot{\cchi}(t')\rangle\,dt\,dt'\\
&=&\!	\int_0^{\Delta t}\!\int_0^{\Delta t}\!
\bold{\mu}\!\cdot\!\langle\bold{F}^{\rm ran}(t)\,
	\bold{F}^{\rm ran}(t')\rangle\!\cdot\!\bold{\mu}\,dt\,dt'\\ &=&\! 
\int_0^{\Delta t}\!\int_0^{\Delta t}\!2T\bold{\gamma}\delta(t-t')\,dt\,dt'
=2T\bold{\mu}\Delta t.\,\,\,\
\eeqar
We note that $\bold{\cal V}T=\bold{\cal D}\cdot\bold{F}^{\rm pot}$
in accordance with the fluctuation-dissipation theorem \cite{EinsteinAdP17}.

\subsection{Direct simulation}

One way to proceed is to solve the Fokker-Planck transport equation (\ref{FPE})
which is a partial differential equation for a time-dependent function
of $N$ variables (in our case $N$=5, see later).
This is a formidable task and we instead perform direct Monte-Carlo 
simulations of the Smoluchowski equation of motion (\ref{EoM})
to generate suitably large samples of individual stochastic shape evolutions.

For this purpose, it is convenient to write the mobility tensor explicitly 
in terms of its eigenvectors $\{\tilde{\cchi}^{(n)}\}$,
normalized such that $\tilde{\cchi}^{(n)}\cdot\tilde{\cchi}^{(n)}$ 
is the eigenvalue $\mu_n$
(which is always positive, as explained above),
\beq\label{XX}
\mu_{ij}\ =\ \sum_n \tilde{\chi}_i^{(n)}\tilde{\chi}_j^{(n)}\ .
\eeq
Assuming now that the current shape is characterized by the value \cchi,
we wish to propagate the shape forward to a slightly later time $t+\Delta t$.
The average shape change is readily obtained from Eq.\ (\ref{dchi}) as
$\delta\overline{\cchi}=\bold{\mu}\cdot\bold{F}^{\rm pot}\Delta t$.
The random contribution to the shape change is most easily sampled in the
eigenframe of the mobility tensor because the increments in each principal 
direction may be sampled separately.
Invoking the eigen representation of $\bold{\mu}$ (\ref{XX}),
we may then obtain the total increment in $\cchi$ 
accumulated in the course of the small time interval $\Delta t$,
\beq\label{deltachi}
\delta\cchi\ =\ \sum_n \tilde{\cchi}^{(n)}
\left[\Delta t\,\tilde{\cchi}^{(n)}\!\cdot\bold{F}^{\rm pot}
+ \sqrt{2T\Delta t}\,\xi_n\right]\ ,
\eeq
where $\{\xi_n\}$ are random numbers 
sampled from a standard normal distribution having zero mean and unit variance.
This propagation procedure is easily implemented,
once the mobility tensor \bold{\mu}\ has been diagonalized.
The average of the accumulated change is then
\beq
\delta\bar{\chi}_i\ =\  \Delta t\sum_n 
	\tilde{\chi}^{(n)}_i\,\tilde{\cchi}^{(n)}\!\cdot\bold{F}^{\rm pot}\ ,
\eeq
because $\langle\xi_n\rangle=0$,
and the accumulated correlation $\sigma_{ij}$ becomes
\beq
\sigma_{ij}\ =\ 2T\Delta t\sum_n \tilde{\chi}^{(n)}_i\,\tilde{\chi}^{(n)}_j\ ,
\eeq
because $\langle\xi_n\xi_{n'}\rangle=\delta_{nn'}$.
Both are proportional to $\Delta t$.

It is an important feature of the propagation scheme (\ref{deltachi})
that when the time interval $\Delta t$ is sufficiently small
the generated ensemble of dynamical histories remains unaffected 
by a subdivision of the time interval,
so the numerical solution of the transport problem is robust.
To see this, imagine that the time interval used in Eq.\ (\ref{deltachi})
is subdivided into a number of shorter
intervals, $\{\Delta t^{(m)}\}$, with $\sum_m\Delta t^{(m)}=\Delta t$.
If $\bold{\mu}$, $\bold{F}$, and $T$ remain unchanged during $\Delta t$, 
the resulting combined change in \cchi\
becomes $\delta\cchi = \sum_m\delta\cchi^{(m)}$ with
\beq
\delta\cchi^{(m)} =  \sum_n \tilde{\cchi}^{(n)}
\left[\Delta t^{(m)}\,\tilde{\cchi}^{(n)}\!\cdot\bold{F}^{\rm pot}
+	\sqrt{2T\Delta t^{(m)}}\,\xi_n^{(m)}\right]\ .
\eeq
Recalling that $\langle\xi_n^{(m)}\rangle$ vanishes,
we see that the combined average change remains the same as before,
\beq
\sum_m\langle\delta\cchi^{(m)}\rangle\
=	\sum_n \tilde{\cchi}^{(n)}
(\sum_m\Delta t^{(m)})\,\tilde{\cchi}^{(n)}\cdot\bold{F}^{\rm pot}\ ,
\eeq
as does the accumulated covariance $\sigma_{ij}$,
\beq
\langle(\sum_m\delta\chi_i^{(m)})(\sum_m\delta\chi_j^{(m)})\rangle
= 2T(\sum_m\Delta t^{(m)})\sum_n\tilde{\chi}_i^{(n)}\tilde{\chi}_j^{(n)}
\eeq
because 
$\langle\xi_n^{(m)}\,\xi_{n'}^{(m')}\rangle\,=\!\delta_{mm'}\delta_{nn'}$.
Thus the diffusion process is robust
under changes in the employed time interval $\Delta t$, as it should be.
Of course, this invariance pertains to the {\em distribution} 
of dynamical histories, $P(\cchi;t)$,
whereas any individual trajectory does change when $\Delta t $ is changed.

A related invariance holds when the overall magnitude 
of the dissipation tensor \bold{\gamma} is changed:  
If the elements $\gamma_{ij}(\cchi)$ are all increased 
by the common factor $N(\cchi)$ then the local time evolution 
proceeds at a rate that is $N(\cchi)$ times slower,
but the resulting distribution of shape trajectories,
$P(\cchi;t)$, remains the same.
This convenient feature allows us to arbitrarily rescale the friction
locally to facilitate the numerical treatment,
because we are not here interested in the actual time evolution
but merely in the final distribution of mass divisions.
Such local rescaling of the dissipation tensor
is equivalent to adjusting the local clock rate
which will obviously not affect the outcome of the process
but merely changes how much time is spent at various locations.
For convenience, we shall therefore assume that the eigenvalues 
$\gamma_n(\cchi)$ are one on average, for each particular shape \cchi.

\subsection{Discrete random walk}
\label{discrete}

It is instructive to consider the simple situation when the mobility tensor 
is aligned with the lattice, i.e.\ $\bold{\mu}$ is diagonal,
$\mu_{ij}=\mu_i\delta_{ij}$.
The transport process then reduces to a standard random walk,
i.e.\ Eq.\ (\ref{deltachi}) reduces to
\beq\label{dchii}
\delta\chi_i = \mu_iF_i\Delta t +\sqrt{2T\mu_i\Delta t}\,\xi_i\ .
\eeq
If the potential energy $U$ is known for any value of
the shape parameter \cchi\ 
the local force $\bold{F}$ can be obtained as the corresponding gradient,
$F_i^{\rm pot}=-\del U/\del\chi_i$,
and the transport process (\ref{dchii}) can be readily simulated
to yield an ensemble of evolutions.
Because each increment $\delta\chi_i$ is a real number
(i.e.\ not necessarily an integer),
each evolution is represented by a sequence of shapes 
whose coordinates $\{\chi_i\}$ may take on any fractional value
within the considered parameter domain.

However, the potential energy employed \cite{MollerPRC79}
is available only on a discrete lattice whose 5,315,625 sites are labeled
by the integers $\{IJKLM\}$,
corresponding to integer values of the shape coordinates,
$\{\chi_1,\dots,\chi_5\}$.
By performing a pentalinear interpolation (see Appendix \ref{ijklm}),
we may obtain an approximate representation of the potential energy
for arbitrary (fractional) values of the shape coordinates, $U(\cchi)$,
and the above continuous transport process (\ref{dchii}) may then be simulated.

A simpler but approximate treatment of the above continuous transport problem
consists in performing a random walk on the discrete lattice,
i.e.\ the shape parameters $\{\chi_i\}$ take on only integer values.
This can be conveniently accomplished by means of the standard Metropolis
sampling procedure (see below),
as was originally done in Ref.\ \cite{RandrupPRL106}.
The quality of this approximation depends on the lattice spacing 
and if the spacing is reduced then the discrete random walk 
becomes a  better approximation to the continuous transport process.

\begin{figure}[t]	
\includegraphics[width=3.4in]{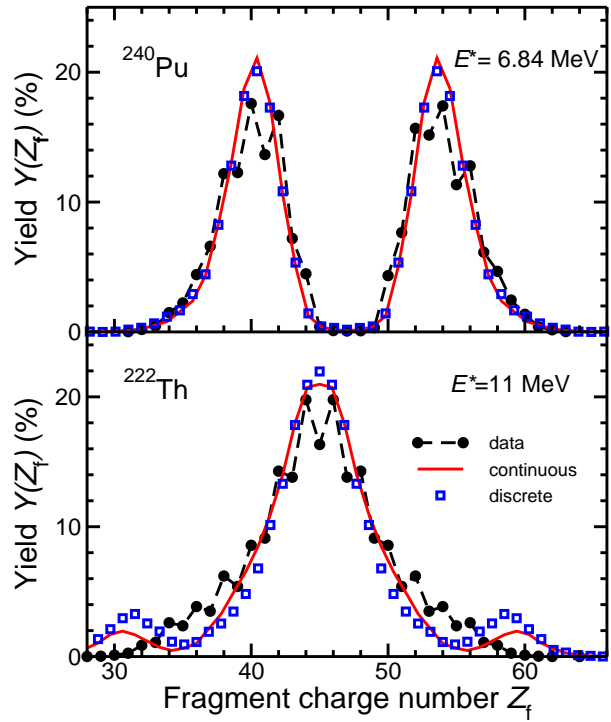}	
\caption{(Color online.)
The effect of the discretization of the shape space 
on the calculated fragment charge distribution is illustrated by comparing 
the results of the Metropolis walk introduced in Ref.\ \cite{RandrupPRL106}
with those of the corresponding continuous process;
the experimental data \cite{ENDF,SchmidtNPA665} are also shown.
Throughout, as in Ref \cite{RandrupPRL106},
the calculated mass yields $Y(A_{\rm f})$
have been transformed to charge yields 
by a simple scaling, $Y(Z_{\rm f})=(A_0/Z_0)Y(A_{\rm f})$,
where $Z_0$ and $A_0$ are the charge and mass numbers 
of the fissioning nucleus.
}\label{f:discrete}
\end{figure}		

To understand how the continuous random walk (\ref{dchii})
can be approximated by a Metropolis procedure,
we first note that when the mobility tensor is diagonal,
each lattice direction $i$ is an eigen direction and can,
therefore, be sampled independently.
In the discrete treatment, 
the step size in the direction $i$ is fixed by the lattice spacing $\Delta_i$,
and we need to know the probabilities for taking a forward or backward step
during a brief time interval $\Delta t$,
$P_\pm^{(i)}=\nu_\pm^{(i)}\Delta t$.
The associated Fokker-Planck transport coefficients,
which express the rate at which the mean location changes
and half the rate at which the variance in the location grows, 
are therefore given by
\beqar
{\cal V}_i &=&~ (\nu_+^{(i)}-\nu_-^{(i)})\Delta_i=\mu_iF_i\ ,\\
{\cal D}_i &=& \half(\nu_+^{(i)}+\nu_-^{(i)})\Delta_i^2=\mu_iT\ ,
\eeqar
where $F_i$ is the force in the lattice direction $i$,
and $\mu_i=1/\gamma_i$ is the mobility in that direction;
they are independent of the lattice spacing $\Delta_i$.
These relations can be readily solved for the rates,
\beq
\nu_\pm^{(i)}\ =\ {\mu_i\over\Delta_i^2}[T\pm\half F_i\Delta_i]\
\approx\ {\mu_i\over\Delta_i^2}[T\mp\half\Delta U_i]\ ,
\eeq
where $\Delta U_i\equiv -F_i\Delta_i$ is the change in the potential 
associated with an increase of $\chi_i$ by $\Delta_i$.
It then follows that
$P_+^{(i)}/P_-^{(i)}\approx\exp(-\Delta U_i/T)$,
which is precisely what characterizes the Metropolis procedure,
as we now show.

In the Metropolis procedure \cite{MetropolisJCP26}, 
a proposed step is always accepted 
if the associated energy change $\Delta U$ is negative,
whereas it is accepted only with the probability $\exp(-\Delta U/T)$ otherwise.
The probabilities for accepting the reverse step are then
$\exp(\Delta U/T)$ or unity, respectively.
So the ratio between the forward and backward acceptance probabilities
is $P_+/P_-=\exp(-\Delta U/T)$.
Thus a continuous random walk of the type (\ref{dchii})
can be treated approximately by means of a discrete random walk
based on the Metropolis procedure.

\begin{figure}[t]	
\includegraphics[width=3.4in]{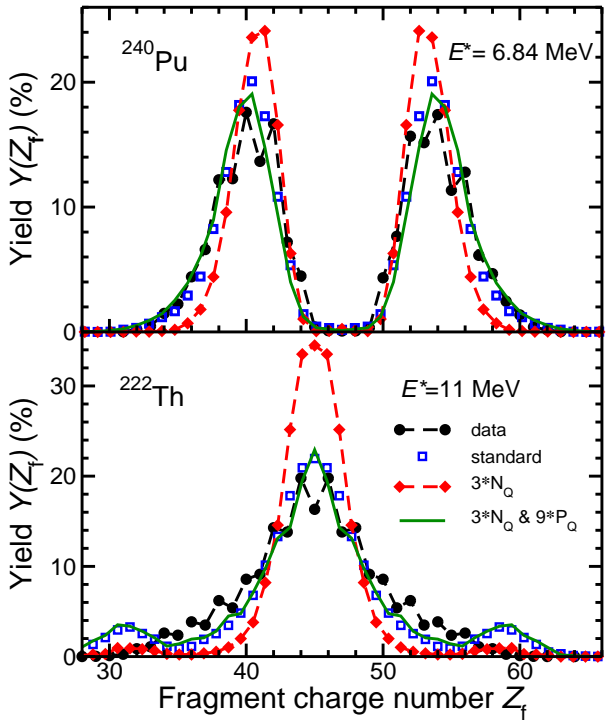}	
\caption{(Color online.)
The effect of changing the lattice spacing for the Metropolis walk: 
The standard lattice \cite{MollerPRC79}
is modified in the $Q_2$ direction by introducing two additional lattice sites 
between the  original lattice sites, thus tripling the density of lattice sites
in that direction;
subsequently the likelihood for considering the $Q_2$ direction
as a candidate for the next step is enhanced by a factor of nine (solid curve)
which is seen to compensate exactly for the reduced lattice spacing.
Also shown are the experimental data \cite{ENDF,SchmidtNPA665}.
}\label{f:mul}
\end{figure}		

The treatment reported in Ref.\ \cite{RandrupPRL106} employed such a
discrete random walk on the 5D lattice of potential energies.
A bias potential of the form $V_{\rm bias}(Q_2)=V_0Q_2^2/Q_0^2$ was added
to disfavor compact shapes and thus help to guide the walk towards
the scission region, thereby speeding up the calculation;
the strength used was checked to be sufficiently small
that no further reduction would affect the mass yields.
Thus the only physical parameter was the critical neck radius, $c_0$,
at which it was assumed that no further change in mass asymmetry would occur;
typically the calculated mass yields were relatively insensitive to $c_0$
in the range of 2-3~fm and $c_0=2.5~{\rm fm}$ was adopted as the standard
value in Ref.\ \cite{RandrupPRL106}.

We are now in a position to ascertain the inaccuracies that are
inherent in such an approach due to the specific lattice employed
in Ref.\ \cite{MollerPRC79}.
We first examine the effect of the finite magnitude of the lattice spacing
by comparing the results of the Metropolis walk with those
of the corresponding continuous process.
This is illustrated in Fig.\ \ref{f:discrete} for two typical cases
(the others considered show comparable effects).
The continuous process and its discrete approximation tend to
yield rather similar results for the fragment mass distribution.
But there are some noticeable differences in the region of the asymmetric peak
for the thorium isotopes considered.
Such discrepancies suggest that the results are quite sensitive 
to the underlying potential energy surface 
in that particular region of the shape space
({\em cf.}\ the effect of modifying the Wigner term 
discussed in Sect.\ \ref{Wigner}),
a feature that might help to achieve a better understanding 
of the potential energy.

Secondly, we address the important role played by the relative size 
of the site spacings in the different lattice directions.
The Metropolis walks carried out in Ref.\ \cite{RandrupPRL106} 
treated all the five lattice directions equally which amounts to 
implicitly assuming that the underlying mobility tensor is isotropic,
i.e.\ the mobilities are the same in all the lattice directions,
$\bold{\mu}\sim\bold{I}$.
Each lattice direction has then an even chance
 for being considered as a candidate for the next Metropolis step.
However, as the above analysis brings out,
a change in the lattice spacing in one direction
modifies the effective mobility in that particular direction:
with an increased density of lattice sites, it takes a larger number 
of elementary Metropolis steps to go a given distance.
Therefore, if the density of lattice sites is increased by a factor of $N$
in the direction $i$ then the corresponding mobility coefficient
is decreased by the factor $N^2$, $\mu_i\to\mu_i/N^2$.
Conversely, to ensure that the evolution of the transport process 
remains unaffected by the increased density of lattice sites,
the likelihood that the affected direction
is being considered as a candidate for the next step
must be increased by that factor.
For example, if the density of lattice sites in a particular direction 
is doubled, $N=2$, then the likelihood for considering that direction 
should be increased from \mbox{$1\over5$}
to $2^2/(4+4)=\mbox{$1\over2$}$ to achieve the same evolution.

These features are illustrated in Fig.\ \ref{f:mul}
for the same two cases that were shown in Fig.\ \ref{f:discrete}.
We consider the effect of changing the lattice spacing in the $I$ direction
which corresponds to the overall elongation 
as quantified by the quadrupole moment $Q_2$.
When the lattice spacing in the $I$ direction is decreased by a factor of three
and the Metropolis walk is repeated with no other change,
then the resulting fragment mass distribution is affected quantitatively,
becoming noticeably narrower.
However, as suggested by the analysis above,
when the corresponding mobility is also increased by a factor of nine
(by favoring the consideration of the $I$ direction correspondingly)
then the mass distribution reverts to its original form.

The relative lattice spacings are thus
intimately related to the anisotropy of the effective mobility tensor.
This basic feature makes the Metropolis walks performed in 
Ref.\ \cite{RandrupPRL106} seem somewhat arbitrary, because a different
choice of lattice spacing, without any compensating change in 
the mobility coefficient, would generally lead to a different final result.
It is therefore important to employ a mobility tensor based on a
physically plausible form of the dissipation
and to investigate the sensitivity of the calculated mass distributions
to that specific structure.
We now turn to this central issue.

\section{Inclusion of dissipation}
\label{mu}

As discussed above, the simple random walk introduced 
in Ref.\ \cite{RandrupPRL106} is most easily justified 
if the dissipation tensor is isotropic in the employed lattice variables.
Since such an idealized scenario is not likely to be realistic,
we wish to study the effect of using a more plausible dissipation tensor,
which is generally anisotropic and has a structure 
that varies from one shape to another.

The potential energy of deformation, $U(\cchi)$,
was calculated \cite{MollerPRC79}
on a lattice of shapes introduced by Nix \cite{UCRL,NixNPA130}.
As described in Appendix \ref{3QS},
each shape is composed of three smoothly joined quadratic surfaces.
These 3QS shapes are characterized by the parameters $\bold{q}=\{q_\mu\}$.
While these are in principle known functions of the lattice shape variables 
$\cchi=\{\chi_i\}$, they are readily available only at the discrete lattice
sites $\cchi=(I,J,K,L,M)$.

The dissipation tensor can be determined from
the rate of energy dissipation $\dot{Q}$,
which is a positive definite quadratic form in the shape velocities,
\beq
\dot{Q}\ =\	\sum_{ij} \dot{\chi}_i\,\gamma_{ij}(\cchi)\,\dot{\chi}_j\
=\	\sum_{\mu\nu} \dot{q}_\mu{\sf g}_{\mu\nu}(\bold{q})\dot{q}_\nu\ .
\eeq
The first term expresses $\dot{Q}$ in terms of the lattice variables 
$\{\chi_i\}$, while the second term uses the 3QS parameters $\{q_\mu\}$
so ${\sf g}_{\mu\nu}(\bold{q})$ is the friction tensor 
with respect to these variables (see Appendix \ref{3QS}).
Once ${\sf g}_{\mu\nu}(\bold{q})$ has been calculated (see below),
we may obtain the required dissipation tensor $\bold{\gamma}(\cchi)$
by the appropriate transformation,
\beq\label{gammaij}
\gamma_{ij}(\cchi)\ =\ \sum_{\mu\nu}\,{\del q_\mu\over\del\chi_i}\,
{\sf g}_{\mu\nu}(\bold{q})\,{\del q_\nu\over\del\chi_j}\ .
\eeq
We wish to determine $\bold{\gamma}(\cchi)$ at the various lattice sites,
at which the parameters $\{\chi_i\}$ have integer values
and we approximate the derivatives $\del q_\nu/\del\chi_n$
in terms of differences between the values of $q_\nu$ at the neighboring sites.
Although this is a relatively rough approximation
because the dependence of $\bold{q}$ on \cchi\ is generally not linear,
it will sufffice for our present explorative purposes.

\begin{figure}[tbh]	
\includegraphics[width=3.4in]{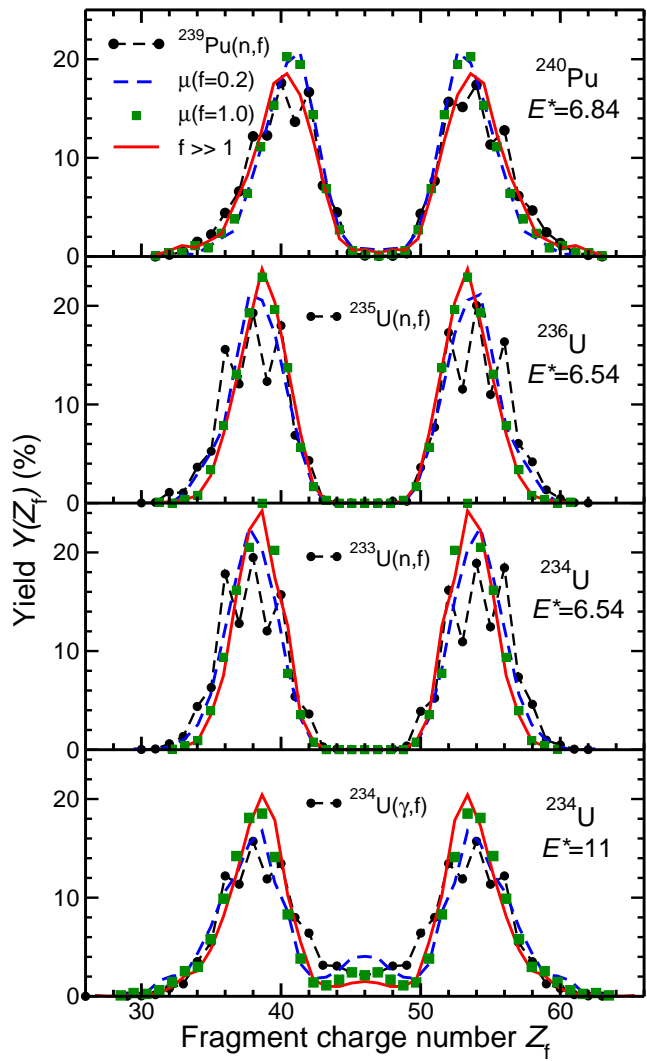}	
\caption{
The charge yields for neutron-induced fission of $^{240}$Pu and 
$^{236,234}$U calculated with increasingly isotropic mobility tensors 
as obtained by using Eq.\ (\ref{muf}) with $f=0.2, 1, \infty$ 
(the latter is fully isotropic and has been indicated as $f\gg1$),
together with the experimental data \cite{ENDF,SchmidtNPA693}.
Those in the top three panels are for (n$_{\rm th}$,f) reactions \cite{ENDF},
while the data in the bottom panel is for ($\gamma$,f) reactions 
leading to $E^*\!\!\approx8-14$~MeV;
they include contamination from multi-chance fission 
\cite{SchmidtNPA665}.
}\label{f:PuUF}
\end{figure}		

In order to calculate ${\sf g}_{\mu\nu}(\bold{q})$,
we thus need to know the dissipation rate $\dot{Q}(\bold{q},\dot{\bold{q}})$.
For that we employ here the ``wall formula'' for
the one-body dissipation mechanism \cite{BlockiAP113}.
The underlying mechanism is the reflection of individual nucleons
off the moving surface which generates a dissipative force
that is rather strong due to the nucleonic Fermi motion.
Because the individual nucleons reach the moving surface at random times,
and from random directions, the associated force on the surface is stochastic,
in accordance with the fluctuation-dissipation theorem \cite{EinsteinAdP17}.
While this idealization may not give a quantitative account
of the actual dissipation rate in a real fissioning nucleus,
it does serve well as a means to provide
us with a mobility tensor that has a quasi-realistic structure.

\begin{figure}[t]	
\includegraphics[width=3.4in]{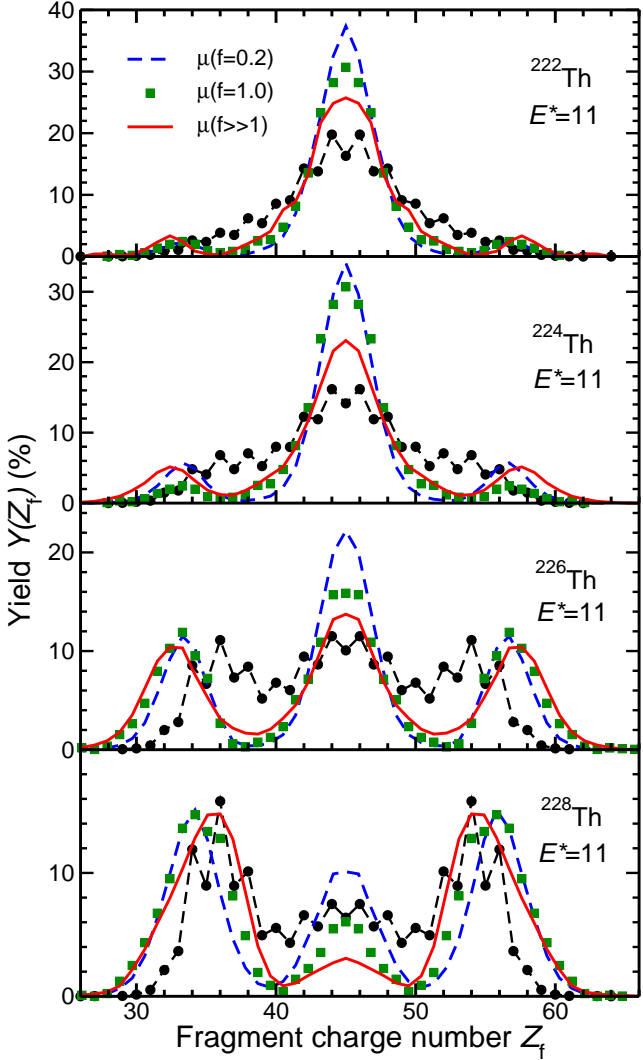}	
\caption{
The charge yield for thorium isotopes
calculated with increasingly isotropic mobility tensors 
as obtained by using Eq.\ (\ref{muf}) with $f=0.2, 1, \infty$ 
(the latter being fully isotropic),
together with the experimental data obtained with ($\gamma$,f) reactions 
leading to $E^*\!\!\approx8-14$~MeV;
they include contamination from multi-chance fission \cite{SchmidtNPA665}.
}\label{f:ThF}
\end{figure}		

In its simplest form,
the one-body dissipation rate in a deforming nucleus is given by 
the simple wall formula, $\dot{Q}=m\rho_0\bar{v}\oint \dot{n}(a)^2d^2a$,
where $m$, $\rho_0$, and $\bar{v}$ 
are the nucleonic mass, density, and mean speed in the interior,
while $\dot{n}(a)$ is the normal surface velocity 
at the location $a$ \cite{BlockiAP113,KooninNPA289}.
It is elementary to show \cite{SierkPRC21}
that the elements of the associated dissipation tensor are given by
\beq\label{gamma}
{\sf g}_{\mu\nu} 
=	\mbox{$\pi\over2$}\rho_0\bar{v}\!\int
	{\del\rho^2(z)\over\del q_\mu}
	{\del\rho^2(z)\over\del q_\nu}
\left[\rho^2+\quart({\del\rho^2\over\del z})^2\right]^{-{1\over2}}\!\!\!\!\!dz,
\eeq
where $\rho(z;\bold{q})$ is the transverse extension of the nucleus
at the position $z$ along the symmetry axis, for the specified values
of the 3QS parameters $\bold{q}$.
The required quantities, namely $\rho^2(z)$ and its derivatives 
with respect to both $z$ and the shape variables $\{q_\nu\}$,
can be expressed analytically for the 3QS shape family \cite{UCRL}
and so it is possible to calculate the elements of $\bold{\sf g}$
for each specified shape.

However, as explained in Appendix \ref{3QS}, certain elements of the 
dissipation tensor $\bold{\sf g}$ may occasionally tend to zero (this happens 
when one of the three quadratic surfaces covers a negligible $z$ interval 
so that this shrinking section ceases to contribute to the dissipation).
Of course, the corresponding derivatives $\del q_\mu/\del\chi_i$
diverge at the same time so the resulting elements of $\bold{\gamma}$
remain well behaved.  But, because it is impractical, for the time being, 
to calculate those derivatives with high accuracy,
the calculation of $\bold{\gamma}$ is correspondingly inaccurate,
with some eigenvalues occasionally becoming unrealistically small.

Fortunately, our main purpose here is merely to study 
the sensitivity of the calculated fragment mass distributions 
to the structure of the dissipation tensor
and we therefore perform the following isotropization procedure.
If $\bold{\gamma}$ is the original tensor,
calculated as described above but renormalized so that its five eigenvalues
$\{\gamma_n\}$ are one on average, i.e.\ $\sum_n\gamma_n=5$,
then we define a more isotropic tensor $\tilde{\bold{\gamma}}$
by modifying the eigenvalues,
\beq\label{muf}
\tilde{\gamma}_n^{(f)}\ \equiv\ [\gamma_n+f]/(1+f)\ ,
\eeq 
where the isotropization coefficient $f$ is a positive number.
We see that the original friction tensor is recovered
when $f$ tends to zero, 
while it approaches isotropy when $f$ grows large.
The corresponding modified mobility tensor $\tilde{\bold{\mu}}(f)$
is then the inverse of $\tilde{\bold{\gamma}}(f)$.

The sensitivity of the calculated charge yields to the degree of structure
in the mobility tensor is illustrated in Figs.\ \ref{f:PuUF}--\ref{f:ThF}
for the cases presented in Ref.\ \cite{RandrupPRL106}.
In addition to the experimental data, which are shown for reference,
each plot shows the result of three different mobility scenarios:
the idealized scenario (labeled $f\!\gg\!1$)
where the mobility tensor is isotropic, an intermediate scenario  ($f=1$)
in which the dissipation tensor is the average of the one calculated with 
the wall formula as described above and the corresponding directional average,
and a more structured scenario ($f=0.2$) 
in which the isotropic admixture is only $20\,\%$.

For the three neutron-induced cases, $^{239}$Pu(n,f) and $^{235,233}$U(n,f),
the change in $Y(Z_{\rm f})$ is very small
as one form of the mobility tensor is replaced by another,
the most noticeable change being a slight narrowing of the asymmetric peaks
for $^{233}$U(n,f).
The photon-induced reactions, which are all calculated for $E^*$=11~MeV,
display a somewhat larger sensitivity.
Generally, as the idealized isotropic mobility tensor grows more anisotropic
there is a tendency for the symmetric yield component to become more
prominent, but the quantitative effect is relatively modest.
It is particularly noteworthy that the evolution from a symmetric yield
for $^{222}$Th to a mixed but predominantly asymmetric yield for $^{228}$Th
remains present in all the scenarios.
When comparing with these data, it should be kept in mind that they
represent a range of excitations ($E^*\approx8-14$~MeV) and
also are contaminated by second- and third-chance fission.

\section{Discussion}
\label{examples}

We now discuss a number of interesting aspects
that can be elucidated with the present treatment.

The choice of shape degrees of freedom made in Ref.\ \cite{MollerPRC79},
and of the specific 5D shape lattice used 
for the calculations of the potential energy,
was guided in large part by physical intuition
(using the somewhat vague but reasonable criterion that the typical
energy change between neighboring sites should be of comparable magnitude).
Our present studies suggest that this lattice of nuclear shapes 
was indeed well chosen,
because the simple Metropolis walks provide mass yields that are
changed only moderately when a more refined treatment is made.
Because the actual mobility tensor is not yet well known,
it would seem prudent to employ a range of mobility scenarios.
The spread among the results might then
be taken as a rough indication of the uncertainty in the prediction.

Furthermore, on the basis of our studies,
it appears that the Metropolis walk,
which is significantly faster than the Smoluchowski simulation
(by 1-2 orders of magnitude), offers a very quick and easy means 
for obtaining practically useful fission-fragment mass distributions.

\subsection{Shape family}
\label{D}

\begin{figure}[b]	
\includegraphics[width=3.4in]{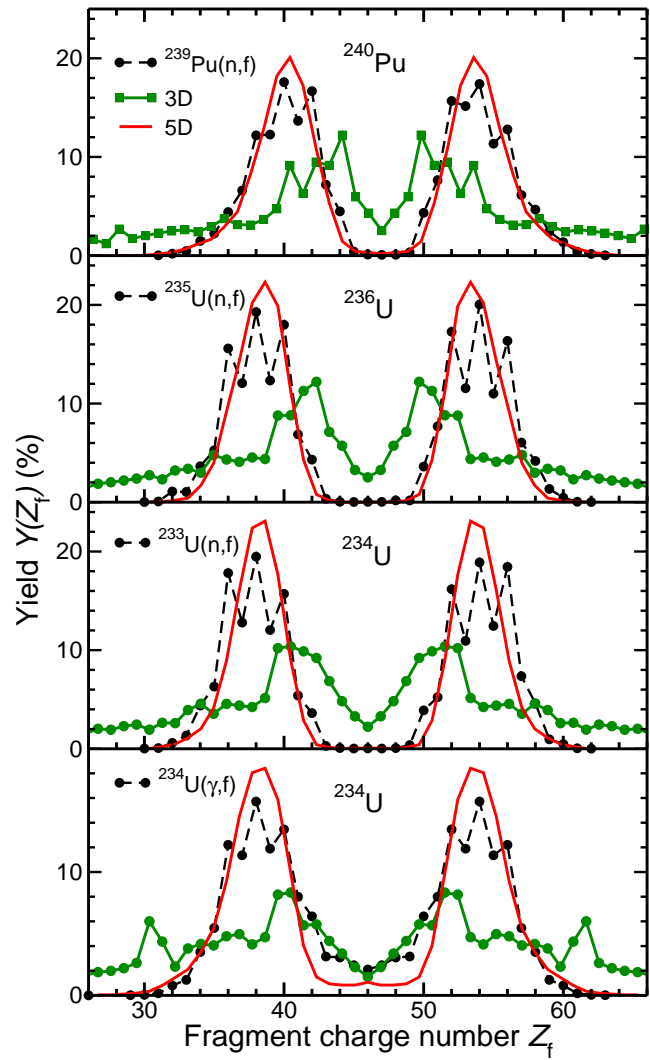}	
\caption{(Color online.)
Charge yields resulting from Metropolis walks on the
5D potential-energy surfaces associated with the 3QS shape family,
together with the corresponding results obtained with 3D surfaces 
generated by minimizing the 5D surfaces w.r.t.\ the individual 
fragment deformations, $\epsilon_{\rm f1}$ and $\epsilon_{\rm f2}$.
Also shown are the experimental data \cite{ENDF,SchmidtNPA665}.
}\label{f:DPuU}
\end{figure}		

We start by illustrating the importance of employing a shape family
that has a sufficient degree of flexibility.
For that purpose, 
we construct three-dimensional potential-energy surfaces
by minimizing the full five-dimensional 3QS surfaces with respect to the
deformations of the two spheroids, $\epsilon_{\rm f1}$ and $\epsilon_{\rm f2}$
(corresponding to the lattice indices $K$ and $L$).
Thus the shapes in the lower-dimensional space are characterized by only
their overall elongation (represented by the lattice index $I$), 
their constriction (represented by the lattice index $J$), 
and the degree of reflection asymmetry (represented by the lattice index $M$). 

\begin{figure}[t]	
\includegraphics[width=3.4in]{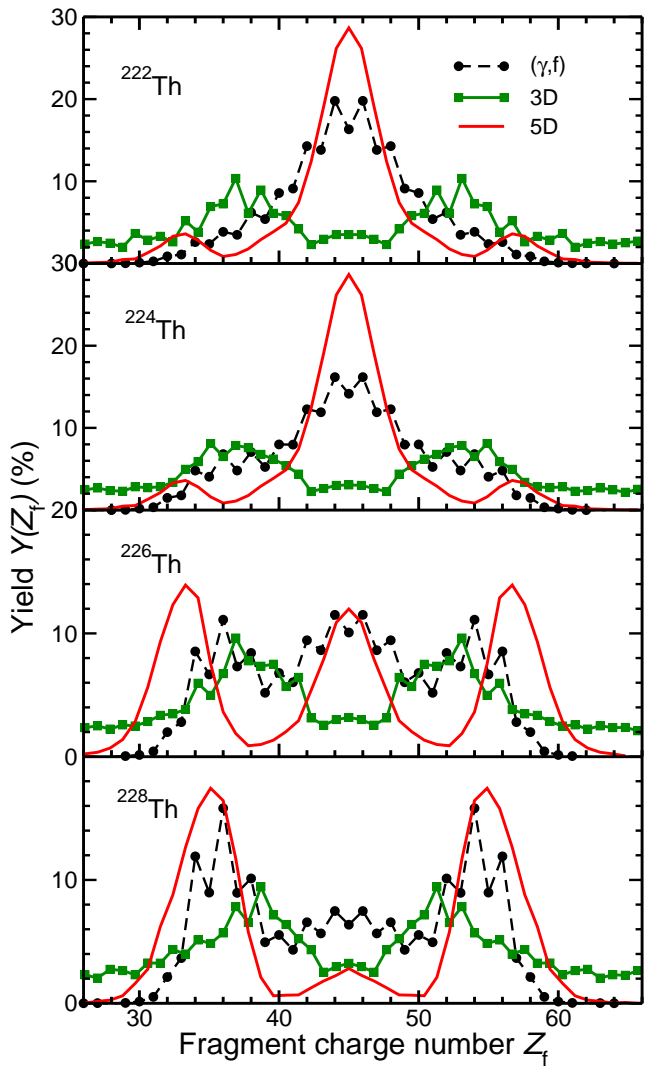}	
\caption{(Color online.)
Similar to Fig.\ \ref{f:DPuU}, 
but for fission of the isotopes $^{222,224,226,228}$Th;
the data are from Ref.\ \cite{SchmidtNPA665}.
}\label{f:DTh}
\end{figure}		

Figures \ref{f:DPuU} and \ref{f:DTh} show the resulting charge distributions 
for the cases presented in Ref.\ \cite{RandrupPRL106},
together with the experimental data 
and our standard results based on the full 5D 3QS shape family.
We see that the although the 3D calculations occasionally reproduce
the qualitative appearance of $Y(Z_{\rm f})$ reasonably well,
the reproduction of the experimental data is generally
far inferior to the results obtained with the 5D shape family.

These examples demonstrate that
it is important to employ a sufficiently rich family of shapes 
when seeking to describe the shape evolution during nuclear fission.
To obtain a reasonably flexible family of shapes,
at least five shape degrees of freedom appear to be required,
namely overall elongation, constriction, reflection asymmetry,
and deformations of the individual prefragments.

\subsection{Saddle shape}
\label{Th222}

\begin{figure}[tbh]	
\includegraphics[angle=0,width=2.0in]{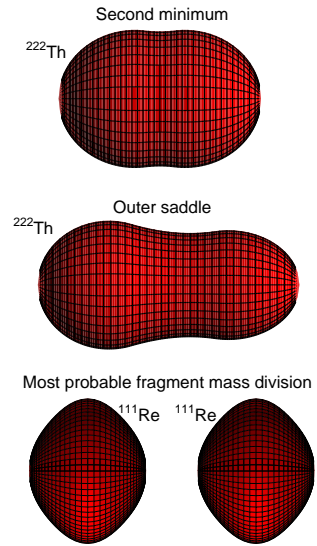}	
\caption{(Color online.)
Three shapes relevant for fission of $^{222}$Th:
the outer potential-energy minimum (which is reflection symmetric)
where the fission isomeric state resides ({\em top}),
the outer saddle which is asymmetric ({\em middle}), 
and the most probable fragment division which is symmetric ({\em bottom}).
}\label{f:shapes}
\end{figure}		

Because our calculational method emulates the actual equilibration process
it is possible to gain insight into the shape evolution during fission.
A particularly instructive case is presented by $^{222}$Th:
Our calculations yield a symmetric mass distribution,
in agreement with the experimental data,
even though they are based on a potential-energy surface
whose fission saddle point corresponds to a nuclear shape 
that is reflection asymmetric.

The most relevant shapes are shown in Fig.\ \ref{f:shapes}.
As the shape evolves from that of the ground state, it tends to pass
near by the second (isomeric) minimum and the nucleus will typically
remain trapped in that minimum for quite some time before escaping,
either back to the ground-state region or towards scission.
(For that reason, we usually start our calculations at the second minimum,
which reduces the required computational effort very significantly;
we have of course checked that this does not alter the results.)
As the figure shows, the outer minimum of $^{222}$Th is reflection symmetric
while the outer saddle lies in a region of significant asymmetry.
Nevertheless, the shapes evolve in such a manner that the final
fragment mass distribution is centered around symmetry.

\begin{figure}[b]	
\includegraphics[angle=-90,width=3.12in]{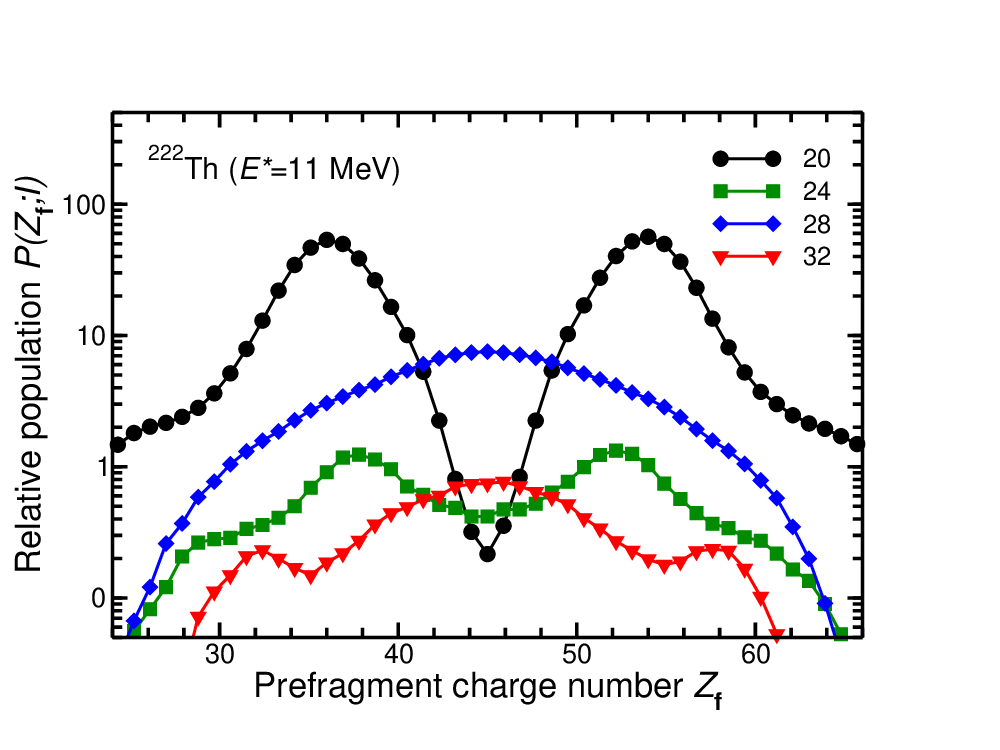}	
\caption{(Color online.)
The evolution of the mass-asymmetry distribution from the region
of the second saddle ($I=20$) towards scission is illustrated by
the charge distribution $P(Z_{\rm f};I)$ for increasing values of $I$
(the lattice index giving the quadrupole moment $Q_2$ of the nuclear shape),
as obtained for an ensemble of 10,000 Metropolis walks.
}\label{f:PQ}
\end{figure}		

More detailed insight into this evolution can be obtained by considering
the charge-asymmetry distribution at specified values of the lattice index $I$
which is a measure of the overall quadrupole moment of the fissioning shape,
$P(Z_{\rm f};I)$.
This conditional distribution is shown in Fig.\ \ref{f:PQ} for increasing
elongations, starting at the value associated with the second saddle point,
$I_{\rm saddle}=20$.
Because it is energetically favorable for the system to traverse the
barrier with an asymmetry close to that of the saddle shape,
$P(Z_{\rm f};I_{\rm saddle})$ is concentrated around that asymmetry.
However, beyond the saddle the preferred asymmetry 
tends to become smaller ($I=24$)
and the asymmetry eventually becomes peaked at symmetry ($I=28$).
As the shape evolves further towards scission ($I=32$),
$P(Z_{\rm f};I)$ developes a minor asymmetric component
that presumably reflects the detailed (and possibly inaccurate) structure
of the potential-energy surface in the scission region
(see the discussion of the Wigner term in Sect.\ \ref{Wigner}).

This result invalidates the commonly made assumption
(see {\em e.g.}\ Refs.\ 
\cite{MollerNPA192,BolsterliPRC5,BenlliureNPA628,SchmidtNPA693})
that the character of the mass distribution, whether symmetric or asymmetric,
is determined by the character of the saddle shape.
In contrast, analyzes of the type illustrated in Fig.\ \ref{f:PQ}
suggest that the structure of the potential-energy landscape 
in the entire region between the isomeric minimum and scission
plays a role in determining  the fragment mass distribution.
Obviously, any plausible model of the mass yields must take this into account.

\subsection{Wigner term}
\label{Wigner}

The generality of our treatment makes it possible to exploit
the remaining differences between calculated results and experimental data
to gain novel insight into aspects of the fission process
that would not otherwise be readily accessible.
As an example, we consider here the shape dependence of the
Wigner term in the macroscopic nuclear energy
\cite{myers77:a,moller89:a,moller94:b,myers96:a,myers97:a,MollerPRL92}.

\begin{figure}[tbh]	
\includegraphics[width=3.4in]{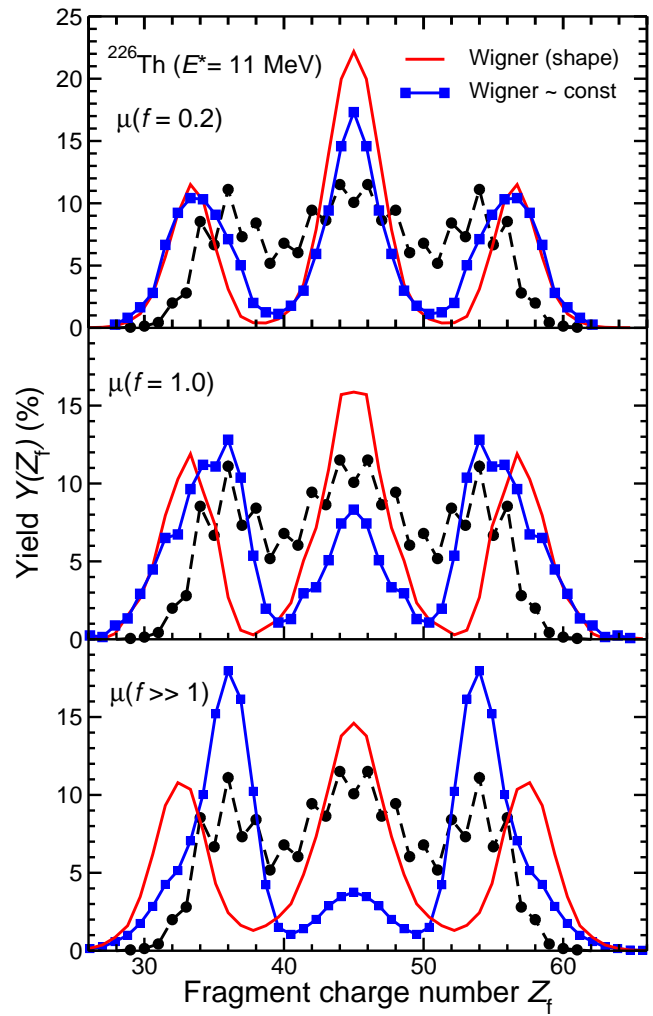}	
\caption{(Color online.) The sensitivity of the charge yields
to the shape dependence of the Wigner term is illustrated for $^{226}$Th,
using either the standard potential-energy surface \cite{MollerPRC79}, 
in which the Wigner term changes gradually as the shape evolves,
or a modified energy surface obtained with a Wigner term 
that remains constant until scission occurs;
the three panels show results calculated with increasingly isotropic mobility
tensors as obtained by using Eq.\ (\ref{muf}) with $f=0.2, 1, \infty$;
the experimental data  are also shown \cite{SchmidtNPA665}.
}\label{f:W}
\end{figure}		

As mentioned above, the potential energies of Ref.\ \cite{MollerNat409}
were calculated as the sum of a macroscopic term
and a microscopic (shell) correction, both being shape dependent.
The macroscopic nuclear energy contains the so-called Wigner term
proportional to $|N-Z|$.
Its presence is clearly visible in the systematics of nuclear masses 
which exhibit a $V$ shape near isosymmetry,
but its microscopic origin is still not well understood,
so it is normally modeled as a phenomenological macroscopic term.
In commonly employed models for nuclear masses
\cite{MollerADNDT59,GorielyPRC66}
the Wigner term is usually introduced without a shape dependence.
However, such an ansatz presents a significant (but often ignored) problem 
in fission where a single nucleus is transformed into two separate nuclei,
each having its own Wigner term.
(That each fragment nucleus must give a Wigner contribution
similar to that of the original nucleus is evident
from the phenomenological form of the term  \cite{moller89:a,moller94:b}.)
Thus the Wigner term must double in magnitude during the evolution 
from a single compound system to two separated fragments,
so, consequently, it must depend on the nuclear shape.

The calculations of fission barriers in for example Refs.\
\cite{moller89:a,MollerPRL92,MollerPRC79}
have since 1989 employed a postulated shape dependence
that relates the increase of the Wigner term to
the decrease in the amount of communication between the two
fragments due to the shrinking of the neck.
The resulting Wigner term changes gradually as the nuclear shape evolves 
and affects the potential-energy landscape correspondingly
(an illustrative figure was given in Ref.~\cite{MollerPRL92}).
However, it may be argued that the change in the character of
the fissioning system from mononuclear to dinuclear occurs more abruptly 
than implied by the currently prescribed shape dependence.
To elucidate the importance of the shape dependence of the Wigner term,
we have considered an alternate form in which the term is kept constant
up to the scission configuration, i.e.\ until the neck radius
has shrunk below the specified critical value $c_0$.

In Fig.\ \ref{f:W} we illustrate how such a modification of the calculated
potential-energy surface affects the calculated charge distribution 
of $^{226}$Th, for which the impact is particularly noticeable.
There are two significant differences between the results of the
two sets of calculations.
One is a change in the relative importance of symmetric and asymmetric
fission, with the constant Wigner term leading to more asymmetric
yield.
The other is a shift in the location of the asymmetric yield peaks,
from being several units on the outside of the observed values towards a better
agreement with the data. 
Both effects depend significantly on the structure of the mobility tensor
and could, in principle, be of help in discriminating 
between different models of the dissipation.

\subsection{Level density}
\label{T}

The microscopic part of the potential energy, $\delta U_{\rm sh}(\cchi)$,
is due to the deformation-dependent variations in the single-particle 
level densities in the effective field of the fissioning nucleus.
This structure also affects the dependence of the nuclear temperature $T$
on the excitation energy $E^*$.
Because a change of the local temperature affects the local diffusion rate
but not the drift rate, a change in $T(E^*;\cchi)$ 
may influence the evolution of the shape distribution $P(\cchi)$.

In order to explore the importance of this effect,
we replace the standard Fermi-gas level-density parameter $\tilde{a}_A=A/e_0$
in the formula $E^*=a_AT^2$ by a ``shell-corrected'' generalization 
suggested by Ignatyuk \cite{Ignatyuk},
\beq\label{amicro}
a_A(E^*;\cchi) = \tilde{a}_A\left[
1+(1-\rme^{-E^*/E_{\rm damp}}){\delta U_{\rm sh}(\cchi)\over E^*}\right] ,
\eeq
where $E_{\rm damp}$ characterizes the gradual dissolution
of the shell effects as the excitation energy is increased.
We shall use $e_0=8\,\MeV$ and $E_{\rm damp}=18.5\,\MeV$.
At low excitation, $E^*\to0$,
$a_A$ tends to $\tilde{a}[1+\delta U_{\rm sh}/E_{\rm damp}]$,
while it approaches $\tilde{a}_A$ monotonically as $E^*$ is increased
(when $E^*\gg E_{\rm damp}$ the exponential is close to zero 
and also $\delta U_{\rm sh}/E^*<E_{\rm damp}/E^*\ll1$ 
since $|\delta U_{\rm sh}|<E_{\rm damp}$).

\begin{figure}[t]	
\includegraphics[angle=-90,width=3.4in]{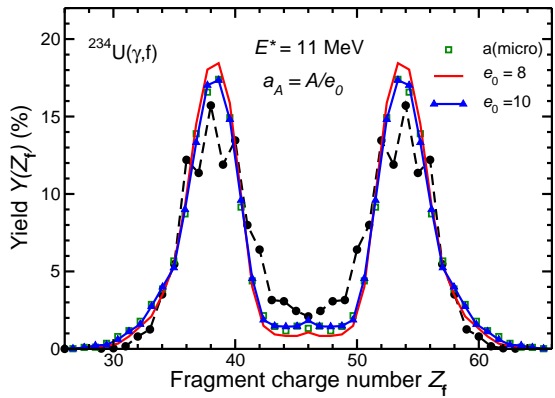}	
\caption{The charge yield for the $^{234}$U($\gamma$,f) reaction
extracted from Metropolis walks using either the macroscopic expression
for the level density, $a_A=A/e_0$, with the standard value $e_0$=8~MeV
and also $e_0$=10~MeV, or the microscopic expression (\ref{amicro});
the experimental data are included \cite{SchmidtNPA665}.
}\label{f:a}
\end{figure}		

We have examined the effect of replacing the macroscopic formula $a_A=A/e_0$ 
by the above microscopic expression (\ref{amicro}).
On the whole, the calculated mass yields are remarkably unaffected
by this change, probably because each random walk visits quite a large
number of lattice sites, so the shell effect, which tends to fluctuate,
largely averages out.
The most noticeable effect occurs for the $^{234}$U($\gamma$,f) reaction
where the use of the microscopic expresison leads to more yield in the
symmetric region (where our standard calculation underpredicts the yield).
This case is shown in Fig.\ \ref{f:a},
where we have also included the result of using the macroscopic formula
with a larger value of $e_0$.
Due to the relation  $E^*=a_AT^2$, this change causes the local temperature
to increase and, as a result, the symmetric yield is increased.
As it turns out the effect of making this increase in $e_0$
in the macroscopic formula is practically identical to the effect of
replacing $a_{\rm macro}$ by $a_{\rm micro}$.

We wish to point out that our present studies do not consider
the effect of pairing which may be included in a simple approximate manner 
by backshifting the excitation energy $E^*(\cchi)$ by the pairing gap 
$\Delta(\cchi)$ when calculating the local temperature $T(\cchi)$.
Such an undertaking would require knowledge of the shape-dependent
pairing gap, $\Delta(\cchi)$.
While this quantity was of course calculated at each lattice site
when the potential-energy surfaces were generated \cite{MollerPRC79},
it was not tabulated seperately, so it is not presently available.
We must therefore leave this interesting issue for future study.

\subsection{Scission model}
\label{sciss}

Finally, we wish to illustrate the importance of the pre-scission 
shape evolution by comparing with the statistical scission model
 \cite{FongPR102,WilkinsPRC14}.
It assumes that the scission configurations,
which are here those having $c(\cchi)=c_0$,
are populated in proportion to $\exp(-U(\cchi)/T(\cchi))$.

Such calculations are, by definition, entirely static in nature
and do not include any dynamical effects. 
While scission models often yield reasonable agreement with the
observed mass distributions, they are not universally successful 
and their failures suggests that mass splits
are sensitive to the pre-scission evolution.
In particular, some scission configurations may not be easily reached
due to the presence of intermediate ridges in the potential-energy landscape.
Such is presumably the case, for example, 
in the recently reported fission of $^{180}$Hg in which 
symmetric splits are strongly favored by the energetics
in the exit channel but are most likely prevented dynamically 
by the presence of a potential-energy ridge,
thus causing the fragment mass distribution to become asymmetric
\cite{AndreyevPRL105,MollerHg}.

\begin{figure}[t]	
\includegraphics[width=3.4in]{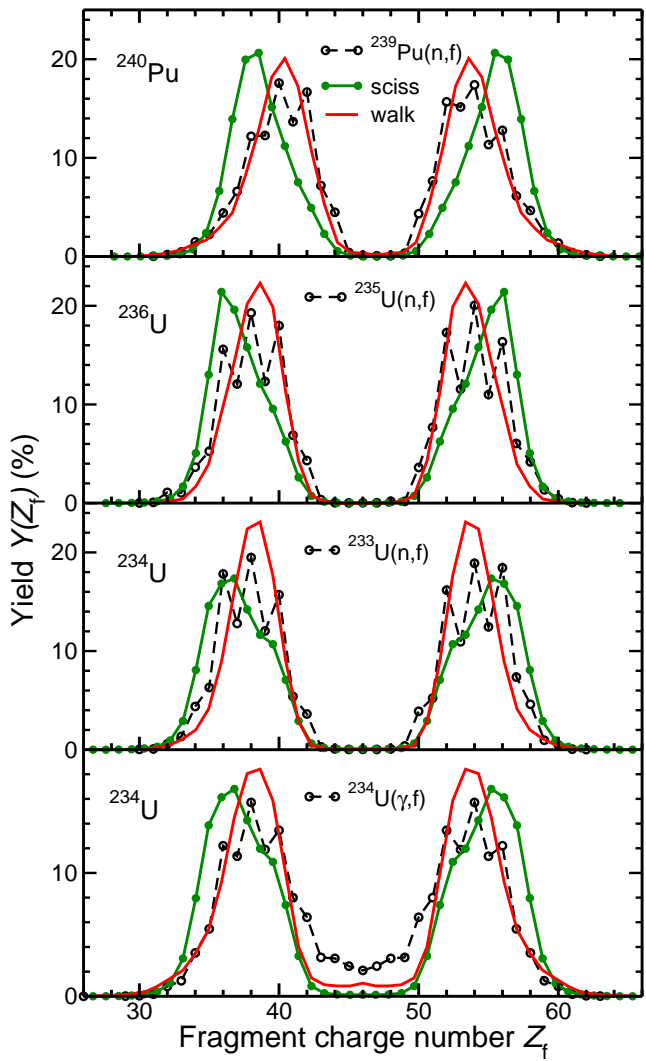}	
\caption{Results of the scission model (see text) for the same plutonium
and uranium isotopes as shown in Fig.\ 6, 
together with the corresponding experimental data and the results of 
the simple Metropolis walk introduced in Ref.\ \cite{RandrupPRL106}.
}\label{f:sciss}
\end{figure}		

But even in cases where there are no such prominent obstacles present in the
potential-energy landscape, the statistical model is often inaccurate.
This is illustrated in Fig.\ \ref{f:sciss} in which its results for
the plutonium and uranium cases are compared with both the experimental data
and the results obtained with the simple Metropolis walk
introduced in Ref.\ \cite{RandrupPRL106}.
The statistical scission model reproduces
the qualitative appearance of $Y(Z_{\rm f})$ reasonably well,
but the quality of the correspondance with the data is inferior 
to that provided by the Metropolis transport calculation.

\section{Concluding remarks}

Generally, calculations of the type discussed here pertain to the
idealized limit of stongly damped motion where the inertia plays no role.
The evolution of the nuclear shape is then akin to Brownian motion.
Whether in fact the shape evolution during fission has such a character is
still an open question, but we believe that this simple limit
provides a useful reference scenario.

In a first exploration of this physical picture,
it was recently shown that simple random walks
on five-dimensional potential-energy surfaces 
lead to remarkably good agreement with experimental data 
on fission-fragment mass distributions \cite{RandrupPRL106}. 
We have examined this method in depth,
studying the importance of a number of effects 
that could be expected to influence the results.

First we generalized the discrete random walk on the fixed lattice sites
where the potential energy is available
to a continuous diffusion process, thus enabling an assessment of the
importance of the finite lattice spacing.
We found that there is typically little difference between the results
of the two treatments, 
although we observed some deviations in certain limited regions.

Such a simple random walk, whether discrete or continuous,
is physically reasonable only if the dissipation tensor happens to be
isotropic in the particular shape variables employed,
which is generally not expected to be the case.
This important feature was illustrated by studying the effect
of inserting additional sites between the standard lattice sites
which reduces the mobility in the affected direction.

The main objective of the present study is therefore to elucidate
how the results of the idealized treatment may change when a more realistic
dissipation tensor is employed.
For this purpose we employed the dissipation tensor suggested by the
simple one-body dissipation wall formula
and introduced an isotropization procedure which allowed us to
examine a continuum of scenarios ranging from perfect isotropy
(corresponding to the idealized random walks discussed above)
to the relatively large degree of anisotropy displayed by the
calculated friction tensors.
Generally we found that the resulting fragment mass distributions
are rather insensitive to the degree of anisotropy,
except, in certain cases only,
for rather extreme anisotropies that are probably unrealistic
and may arise from certain numerical problems.

We then examined a number of additional relevant aspects.
First we demonstrated the importance of using a sufficiently rich
family of shapes by comparing results based on the full 5D potential energy
surfaces with analogous calculations in a reduced 3D deformation space
obtained by constrained minimization.
The importance of using sufficiently flexible shapes was borne out
in particular by the isotope $^{222}$Th whose mass yield is symmetric
even though the outer saddle shape is asymmetric.

As an example of the potential of the present method
for elucidating novel aspects of nuclear properties,
we examined the sensitivity of the calculated mass distribution
to the character of the deformation dependence of the Wigner term
in the macroscopic energy functional.
Specifically, it was demonstrated for $^{226}$Th that 
when the previously employed rather gradual shape dependence was replaced
by a more abrupt transition from the mononucleus to the dinucleus
then the symmetric component of $Y(Z_{\rm f})$ is reduced 
and its asymmetric peaks move towards better agreement 
with the experimental data.

We also examined the effect of taking account of the deformation-dependent
bunchings in the single-particle level densities which affect
the energy dependence of the temperature for a given shape
and thus the local diffusion rate relative to the drift.
The calculated mass yields are relatively unaffected by this refinement, 
except for a favorable shift in the amount of nearly symmetric mass splits,
but it should be kept in mind that the effect of the (presently unavailable) 
shape-dependent paring gap is still left for future study.

Finally, comparisons with mass distributions arising from
a statistical population of the scission configurations
demonstrated the importance of pre-fission shape evolution.
Typically, our transport calculations yielded significantly better
agreement with the experimental data.

For some of the cases considered here, most notably $^{239}{\rm Pu(n,f)}$,
the calculations reproduce the experimentally observed yields very well 
and are rather robust with respect to model variations,
such as changes in the mobility tensor.
Hence, for the comparisions with data to be informative,
it appears to be important to include also cases 
that exhibit larger sensitivity, such as the thorium isotopes.

In conclusion, our studies suggest that the simple Metropolis walk
\cite{RandrupPRL106} on the previously calculated potential-energy 
lattice \cite{MollerPRC79} indeed presents a useful calculational tool
for obtaining the approximate form of fission-fragment mass distributions
for a large range of nuclei.
For more accurate results it is necessary to invoke also
the dissipative features of the shape evolution as represented by the
shape-dependent mobility tensor.
The shape evolution then resembles Brownian motion in an anisotropic 
(and non-uniform) medium.
However, because the dissipation mechanism is not yet as well understood 
as the potential energy,
we propose to make a series of calculations with mobility tensors
that display different degrees of anisotropy and then use the ensuing
spread in the results as an indication of the uncertainty of the
predicted mass yield.
The results obtained in this manner are often remarkably robust.
Consequently, the method may be of practical use for calculating
fission-fragment mass distributions for any of the thousands of nuclei
for which the required 5D potential-energy surface is already available.
~\\

\noindent{\bf Acknowledgments}\\
We thank K.-H.\ Schmidt for providing computer-readable files
of the experimental data in Ref.\ \cite{SchmidtNPA665} and
L.~Bonneau, H.~Goutte, D.C.~Hoffman, A.~Iwamoto, and R.~Vogt 
for helpful discussions; 
T.~Watanabe kindly extracted the (n,f) data from the ENDF/B-VII.0 data base.

This work was supported by the Director, Office of Energy Research,
Office of High Energy and Nuclear Physics,
Nuclear Physics Division of the U.S.\ Department of Energy
under Contract DE-AC02-05CH11231 (JR)
and JUSTIPEN/UT grant DE-FG02-06ER41407 (PM),
and by the National Nuclear Security Administration 
of the U.S.\ Department of
Energy at Los Alamos National Laboratory 
under Contract No.\ DE-AC52-06NA25396 (PM \&\ AJS).\\

\appendix
\section{Lattice interpolation}
\label{ijklm}

The nuclear shapes are characterized by the five shape parameters
$\{\chi_n\}=(\chi_1,\chi_2,\chi_3,\chi_4,\chi_5)$ 
which are collectively denoted by \cchi.
But the potential energy of deformation, $U(\cchi)$, 
is known only on a five-dimensional Cartesian lattice
on which the shape parameters take on the integer values 
$\bold{X}=\{X_n\}=(I,J,K,L,M)$,
whose ranges are given in Ref.\ \cite{MollerPRC79}.
(We use $\bold{X}$ to denote a capital \cchi.)
We describe here how the potential energy for arbitrary $\cchi$ values 
can be obtained by pentalinear interpolation,
i.e.\ an interpolation scheme that yields a function $U(\{\chi_n\})$
that is linear in each of the five variables $\chi_n$ 
inside each elementary hypercube.
[The resulting function is identical to the Taylor expansion around 
the ``lower-left'' hypercube corner, $\bold{X}=(I^-,J^-,K^-,L^-,M^-)$,
keeping only terms of first order in each variable
and approximating all derivatives by the corresponding central differences.]

We assume that the given shape parameter \cchi\ lies within the domain covered
by the lattice and start by identifying the surrounding elementary hypercube.
Its 32 corners are given by the indices $(I^\pm,J^\pm,K^\pm,L^\pm,M^\pm)$,
where  $X_n^-=[\chi_n]$ (i.e.\ the integer part of $\chi_n$)
and $X_n^+=X_n^-+1$ so $X_n^-\leq\chi_n<X_n^+$ for $n=1,\dots,5$.
In a single dimension, the interpolated value would be
\beq
U(\chi)\ =\ U(X^-)(X^+-\chi)\ +\ U(X^+)(\chi-X^-)\ ,
\eeq
so we may readily generalize to five dimensions,
\beqar\label{U5}
&~&U(\cchi) =\! \sum_{ijklm}\!
ijklm\	U(I^{-i}\!,J^{-j}\!,K^{-k}\!,L^{-l}\!,M^{-m})\, \\ \nonumber
&~&\times~
(I^i\!-\chi_1)(J^j\!-\chi_2)(K^k\!-\chi_3)(L^l\!-\chi_4)(M^m\!-\chi_5),
\eeqar
where the summation indices each take on the values $\mp1$.
It is easy to verify that this is indeed correct:
Within the local hypercube (within which \cchi\ is located)
the above expression is linear in each of the five $\chi$ variables
and it yields the correct matching 
because when \cchi\ coincides with a lattice site, 
$\cchi=\bold{X}=(I,J,K,L,M)$, 
then $I^-\!=I$, ..., $M^-\!=M$ so $X_n^--\chi_n=0$ and $X_n^+-\chi_n=1$,
so only the term with $i=+1,\dots,m=+1$ contributes, 
yielding $U(\cchi)=U(I,J,K,L,M)$.

The driving force $\bold{F}=-\grad U(\cchi)$ can be obtained by taking
the derivative of the above expression (\ref{U5}).  Thus
\beqar\label{F5}
&~&F_1(\cchi) = -\sum_{ijklm}\!	\nonumber
ijklm\	U(I^{-i}\!,J^{-j}\!,K^{-k}\!,L^{-l}\!,M^{-m})\\
&~&\times~
(J^j\!-\chi_2)(K^k\!-\chi_3)(L^l\!-\chi_4)(M^m\!-\chi_5),
\eeqar
and analogously for the other four directions.
Thus, within the local hypercube,
the force component $F_n$ does not depend on $\chi_n$,
as is consistent with the fact that
the potential is locally linear in $\chi_n$.

A similar scheme is used to calculate other quantities
for arbitrary shapes, such as the neck radius $c(\cchi)$
and the dissipation tensor $\bold{\gamma}(\cchi)$.

\section{3QS shape family}
\label{3QS}

The three-quadratic-surface shape family introduced by Nix \cite{NixNPA130}
consists of axially symmetric shapes for which the square of the local radial
distance $\rho(z)$ to the surface
is given by three smoothly joined quadratic surfaces,
\addtocounter{equation}{1}\[\label{rho2}
\,\,\,\rho^2(z) = \left\{\begin{array}{ll}
a_1^2-(a_1^2/c_1^2)(z-\ell_1)^2 ,& \ell_1-c_1 \leq z \leq z_1, \\[0.5ex]
a_3^2 -(a_3^2/c_3^2)(z-\ell_3)^2 ,& z_1 \leq z \leq z_2 ,
	\hspace{2.0em} ({\rm \theequation})\\[0.5ex]
a_2^2-(a_2^2/c_2^2)(z-\ell_2)^2 ,& z_2\leq z \leq \ell_2+c_2 .
\end{array}\right.
\]
Thus nine numbers are required to specify the nuclear surface.
Two of these are eliminated due to the
continuity of $\rho(z)$ and its derivative at $z_1$ and $z_2$, and 
the length parameter $u=[\half(a_1^2+a_2^2)]^{1/2}$ governs the overall scale.
The remaining six numbers are then determined by six 
dimensionless shape parameters $\{q_\nu\}$,
\beqar
&~&\hspace{-2.5em}	
	\sigma_1={\ell_2-\ell_1\over u}\ ,\
	\sigma_2={a_3^2\over c_3^2}\ ,\
	\sigma_3=\half\left({a_1^2\over c_1^2}+{a_2^2\over c_2^2}\right)\ ,\\
&~&\hspace{-2.5em}
	\alpha_1={\ell_1+\ell_2\over u}\ ,\
	\alpha_2={a_1^2-a_2^2\over u^2}\ ,\
	\alpha_3={a_1^2\over c_1^2}-{a_2^2\over c_2^2}\ .
\eeqar
Furthermore, if the shapes are required to have a given center of mass,
then the parameter $\alpha_1$ is determined 
once the other five have been specified.
All the shapes in the $IJKLM$ lattice have their center of mass at the origin,
which effectively reduces the six-dimensional $\{q_\nu\}$ space
to the five-dimensional shape space covered by the $IJKLM$ lattice.

For the evaluation of the dissipation tensor $\gamma_{\mu\nu}$
we need the derivatives $\del\rho^2/\del q_\nu$ as well as $\del\rho^2/\del z$
which, though somewhat involved, can be expressed analytically \cite{UCRL}.

The 3QS family includes shapes for which one of the three
segments covers only a negligible $z$ interval.
The contribution from such a segment to the dissipation rate is then also
negligible and, as a result, 
so are the associated elements of the dissipation tensor.
These singularities are numerically inconvenient and must be addressed.

		

			\end{document}